\documentclass[12pt,a4paper]{article}
\usepackage[pdfstartview=FitH,colorlinks=true,linkcolor=black,anchorcolor=black,citecolor=black,urlcolor=black]{hyperref}
\usepackage[english]{babel}
\usepackage{amsmath,amssymb,titling,authblk}
\usepackage{slashed}
\usepackage{amsmath}
\usepackage{amscd}
\usepackage{mathrsfs}
\usepackage[normalem]{ulem}
\usepackage{appendix}
\usepackage{cite}

\ifx\pdfoutput\undefined
\usepackage{graphicx}
\else
\usepackage[pdftex]{graphicx}
\fi

\usepackage{float}\usepackage[font=small]{caption}
\usepackage{subcaption}

\begin{document}
\title{States localized on a boundary of the time-dependent parity-breaking medium}

\author[1]{Oleg O. Novikov\thanks{o.novikov@spbu.ru}}
\author[1,2]{Anna Zakharova\thanks{zakharova.annet@gmail.com}}
\affil[1]{St. Petersburg State University, St. Petersburg, 7/9 Universitetskaya nab., 199034, Russia}
\affil[2]{Petersburg Nuclear Physics Institute named by B.P. Konstantinov of National Research Centre
"Kurchatov Institute", Gatchina, 1, mkr. Orlova roshcha, 188300, Russia}

\maketitle

\begin{abstract}
We consider the massive vector field propagating in the inhomogeneous parity-breaking medium, such as the dense hot hadronic matter with chiral imbalance. The transition between the regions with approximately constant values of the parity-breaking parameter allows for the states localized on such boundary to occur. The adiabatic change of the background introduces either decay or the amplification of the localized states. We also discuss the non-adiabatic destruction of these bound states.

\end{abstract}

\section{Introduction}

It is a well-known feature of the Standard model all the parities $\mathcal{P}$, $\mathcal{C}$, $\mathcal{T}$ and $\mathcal{CP}$ are non-conserved \cite{schwartz2014quantum}. Many models of the new physics predict new sources of the $\mathcal{CP}$ violation. This may affect the physics at the energies much lower than the masses of the new particles through the radiative loop corrections. Thus, the searches of the $\mathcal{CP}$-violating effects in the collider, atomic and molecular experiments\cite{chen2021heavy,khriplovich2012cp, ACME:18, zakharova2022rotating, ourRaOH,zakharova2021rovibrational,petrov2022sensitivity} constitute an important strategy for the exploration of the physics beyond the Standard model.

One of the ways the $\mathcal{CP}$-symmetry may be violated is through the $\theta$-term of the gluon field,
\begin{align}
S=-\frac{\theta}{16\pi^2}\int d^4x\, \mathrm{Tr}\Big[G_{\mu\nu}\tilde{G}^{\mu\nu}\Big],\nonumber\\
\tilde{G}^{\mu\nu}=\frac{1}{2}\epsilon^{\mu\nu\alpha\beta}G_{\mu\nu}
\end{align}
When $\theta$ is constant, this term happens to be a topological charge. Thus, it does not appear in the perturbative processes, however it changes the nonperturbative dynamics. One of the effects is the emergence of the electric dipole moment (EDM) for the neutron \cite{harris1999new}. Its measurement \textcolor{red}{puts} very strong constraints on the value of $\theta$ which constitutes a so-called strong $\mathcal{CP}$-problem \cite{Cheng1988,KimCarosi2010}.

However, even if fundamentally $\theta$ is zero, the strong interaction dynamics may admit the local $\mathcal{P}$ and $\mathcal{CP}$-violation \cite{Kharzeev:1998kz,buckley2000can,kharzeev2006parity,Andrianov:2007kz,Andrianov:2009pm}

Such an effect may occur when the large fluctuations of the gluon field produce metastable configurations with nonzero value of the axial topological charge that is tied through the anomaly to the quark axial charge. The emergence of the parity-violating regions may occur within the fireballs of very hot and very dense hadronic medium approaching a phase transition curve of the QCD phase diagram. Such conditions may be achieved in the heavy-ion collisions at NICA, STAR, PHOENIX, SPS and CBM experiments. The manifestations of the QCD local parity violation include the chiral magnetic effect in the peripheral collisions \cite{kharzeev2008effects,fukushima2008chiral,buividovich2009numerical,abelev2009azimuthal,fukushima2010electric,yamamoto2011lattice,Kharzeev:2013ffa,haque2019measurements} and the excessive production of the dileptons in the central collisions \cite{agakishiev2008study,lapidus2009low,ElectroProbes,andrianov2012dilepton,andrianov2014analysis}. The chiral imbalance environment may also produce exotic decays of mesons \cite{Andrianov:2017hbf,Andrianov:2017ilv}.

The presence of the chiral imbalance background affects the propagation and interaction of the effective photon and meson particles. The finite temperature and axial chemical potential as well as the local nature of the fireball result in the  Lorentz symmetry breaking. This is associated with the local violation of the $\mathcal{CPT}$ invariance.

The fireball of chiral matter exists only for a short time. After the cooldown of the medium and disappearance of the local chiral imbalance, the effective vacuum state transforms into a squeezed state of ordinary photons and mesons \cite{Andrianov:2016swz}.

Similar situation may occur on different scales if the strong $\mathcal{CP}$-problem is solved with help of the axion particle. In that scenario, the constant $\theta$ is promoted to the condensate of the pseudoscalar axion field. The axion field must also have similar $\mathcal{CP}$-violating interaction with the electromagnetic field. Local variations of the condensate of the axion particles may play the role of the Dark matter. In the vicinity of the very dense stars the condensate may be especially high \cite{Andrianov:2015oxa}. The passage of the photons through the spatial boundary of such an axion lump was investigated in \cite{Andrianov:2016swz}. Those results also give a rough idea on the similar passage of the vector mesons through the spatial boundary of the static hadronic medium.

However, the aforementioned papers consider only the very edge of the transition region where $\theta$-parameter behaves linearly. This may be valid only for a very high energy particles that have short wavelengths, and, thus, insensitive to the large scale behavior. In contrast, particles with lower energies should be sensitive to the entire transition area. As will be demonstrated in this paper, the transition between two plateau regions may contain the localized states of the massive vector field. While the model we considered is simplified, this implies that the fireball of the chiral hadronic matter possesses the boundary currents. We also consider the influence of the temporal evolution of the medium which results in the amplification or the decay of the boundary states.

\section{Vector field in the parity breaking medium}

As in \cite{Andrianov:2016swz} we start with the massive electrodynamics model with local parity-violating term,
\begin{align}
S=\int d^4x \Big(-\frac{1}{4}F_{\mu\nu}F^{\mu\nu}-\frac{1}{4}\theta(x)F_{\mu\nu}\tilde{F}^{\mu\nu}\nonumber\\
+\frac{1}{2}m^2A^\mu A_\mu
+A^\mu\partial_\mu B+\frac{1}{2} B^2\Big)
\end{align}
This model is related to the vector dominance model in the presence of the chiral imbalance as described in \cite{Kovalenko:2020ryt}. The auxiliary field $B$ is introduced to employ the St\"{u}ckelberg mechanism for self-consistency of the massive vector field theory \cite{Stueckelberg:1957zz,Batalin:1986fm}.  The equations of motion for this field yield,
\begin{equation}
(\square+m^2)B=0,\quad \partial_\mu A^\mu=B
\end{equation}
where $\square=\partial_\mu\partial^\mu$. From that we can derive the condition,
\begin{equation}
(\square+m^2)(\partial_\mu A^\mu)=0,\label{TransverseCond}
\end{equation}
and omit $B$ from our consideration. The equations of motion for $A^\mu$ take the form,
\begin{equation}
(\square+m^2)A^\mu=\epsilon^{\mu\nu\alpha\beta}(\partial_\nu\theta)(\partial_\alpha A_\beta)
\end{equation}

We will assume that the boundary thickness of the chiral breaking matter region is much smaller than its size. Then we can approximate $\theta$ by a function only of $t$ and $z$. In the components the equations take the form,
\begin{align}
(\square+m^2)A^t=-(\partial_z\theta)\Big[\partial_x A^y-\partial_y A^x\Big],\\
(\square+m^2)A^z=(\partial_t\theta)\Big[\partial_x A^y-\partial_y A^x\Big],\\
(\square+m^2)A^x
=
(\partial_t\theta)\Big[\partial_yA^z-\partial_zA^y\Big]\nonumber\\
+(\partial_z\theta)\Big[\partial_t A^y
+\partial_y A^t\Big],\\
(\square+m^2)A^y
=-(\partial_t\theta)\Big[\partial_xA^z-\partial_zA^x\Big]\nonumber\\
-(\partial_z\theta)\Big[\partial_t A^x+\partial_x A^t\Big]
\end{align}

As the translational invariance in $x$ and $y$ directions is not broken it is useful to rewrite these equations in the momentum representation,
\begin{equation}
A^\mu(t,x,y,z)=\int d^2k A^\mu(t,z)\exp(ik_x x+ik_y y),
\end{equation}
It is also convenient to use the circular components,
\begin{equation}
A_\pm=A^x\pm iA^y,\quad k_\pm=k_x\pm i k_y
\end{equation}
Then the equations take the form,
\begin{align}
(\square+m^2)A^t=-\frac{1}{2}(\partial_z\theta)\Big[k_-A_+-k_+A_-\Big],\\
(\square+m^2)A^z=\frac{1}{2}(\partial_t\theta)\Big[k_-A_+-k_+A_-\Big],\\
(\square+m^2)A_\pm=\pm k_\pm\Big[(\partial_t\theta)A^z+(\partial_z\theta)A^t\Big]\nonumber\\
\pm i\Big[(\partial_t\theta)\partial_z A_\pm-(\partial_z\theta)\partial_tA_\pm\Big]
\end{align}

For the simplicity we will consider the case of the momentum directed perpendicularly to the boundary i.e. $k_\pm=0$, assuming that the results may serve as an approximation for sufficiently low momenta in the $x$ and $y$ directions. Then the $A_\pm$ decouple both from the equations on $A^t$ and $A^z$, and from the condition \eqref{TransverseCond}.

The $(t,z)$ sector does not depend on $\theta$ and corresponds to the longitudinal polarization of the massive photon in the empty space. In contrast the equation for the transverse polarizations looks like,
\begin{equation}
(\partial_t^2-\partial_z^2+m^2)A_\pm=\pm i\Big[(\partial_t\theta)\partial_z A_\pm-(\partial_z\theta)\partial_t A_\pm\Big],\label{TransverseEq}
\end{equation}
As $A_-$ satisfies the same equation as $A_+$ but for the reversed time, we will consider only $A_+$ from now on.

\section{Adiabatically changing parity-violating background}

First, we will consider the equation \eqref{TransverseEq} in the adiabatic regime, when $\theta$ is changing in time slowly compared to the frequency of the wave whereas its spatial variation is sufficiently high compared to the wavelentgth. Then we will assume that the solution behaves as,
\begin{equation}
A_+=a(t,z)\exp\Big(-i\int dt\,\omega(t)\Big)
\end{equation}
where the profile function $a$ and the frequency $\omega(t)$ are slowly varying at the same rate as $\theta$.

Then in the leading approximation the $i(\partial_t\theta)$ term may be neglected. We will denote the solutions in this approximation with subscript $0$,
\begin{equation}
a(t,z)\simeq a_0(t,z),\quad \omega(t)\simeq \omega_0(t)
\end{equation}
and the equation becomes,
\begin{equation}
-\partial_z^2a_0+\Big[m^2-\omega_0^2+\omega_0(\partial_z\theta)\Big]a_0=0,\label{AdiabaticLeading}
\end{equation}
which looks like a Schr\"{o}dinger equation though with the potential proportional to the spectral parameter $\omega_0$.

Consider the boundary between regions of approximately homogeneous $\theta$. Let us consider the following shape of $\theta$,
\begin{align}
\theta(t,z)=\bar{\theta}(t)+\Delta\theta(t)\cdot\tanh\Bigg[\mu(t)\Big(z-z_0(t)\Big)\Bigg],\nonumber\\
\theta\underset{z\rightarrow-\infty}{\longrightarrow}\bar{\theta}-\Delta\theta,\quad \theta\underset{z\rightarrow +\infty}{\longrightarrow}\bar{\theta}+\Delta\theta,\label{ThetaTanh}
\end{align}
where all the functions only slowly depend on time. For this shape of $\theta$ the potential becomes the P\"{o}schl-Teller one \cite{flugge2012practical},
\begin{equation}
\Bigg[-\partial_z^2+m^2-\omega_0^2+\frac{\omega_0\mu\cdot\Delta\theta}{\cosh^2{\mu (z-z_0)}}\Bigg]a_0=0,
\end{equation}
As at large $z$ the potential becomes constant $-\varepsilon\equiv -(\omega_0^2-m^2)$ the potential has the continuous spectrum of the wavelike solutions for $\varepsilon>0$ i.e. $|\omega_0|>m$.

If $\varepsilon<0$ i.e. in the band $|\omega_0|<m$ the bound states may exist. This is indeed true for the case of $\omega\mu\cdot \Delta\theta<0$ when the last term gives a potential well. Then, as well-known for the P\"{o}schl-Teller potential the bound states correspond to,
\begin{equation}
\varepsilon_n=-\mu^2(\lambda-n-1)^2,\quad \lambda(\lambda-1)=\Big|\frac{\omega_0\Delta\theta}{\mu}\Big|
\label{Spectrum}
\end{equation}
where $n$ is integer from $0$ to the largest integer $n_{max}\leq \lambda-1$. As $\varepsilon=\omega_0^2-m^2$ this turns into equation on $\lambda$,
\begin{align}
\lambda^4-2\lambda^3+\Big(1+\Delta\theta^2\Big)\lambda^2-2(n+1)\Delta\theta^2\lambda\nonumber\\
+\Delta\theta^2\Big((n+1)^2-\frac{m^2}{\mu^2}\Big)=0
\end{align}
When this equation has a real solution $\lambda>1$ the bound state exists. From the discussion above it is evident that $n_{max} (n_{max}+1)\leq |m\Delta\theta/\mu|$.

The normalized $n=0$ bound state takes the form,
\begin{align}
a_{0,n=0}=\frac{C}{\cosh^{\lambda_0-1}{\mu (z-z_0)}},\nonumber\\
C^2=\frac{\mu}{\sqrt{\pi}}\frac{\Gamma(\lambda_0-\frac{1}{2})}{\Gamma(\lambda_0-1)}.\label{ZeroMode}
\end{align}

\begin{figure}[h]
\centering
  \includegraphics[width=0.45\textwidth]{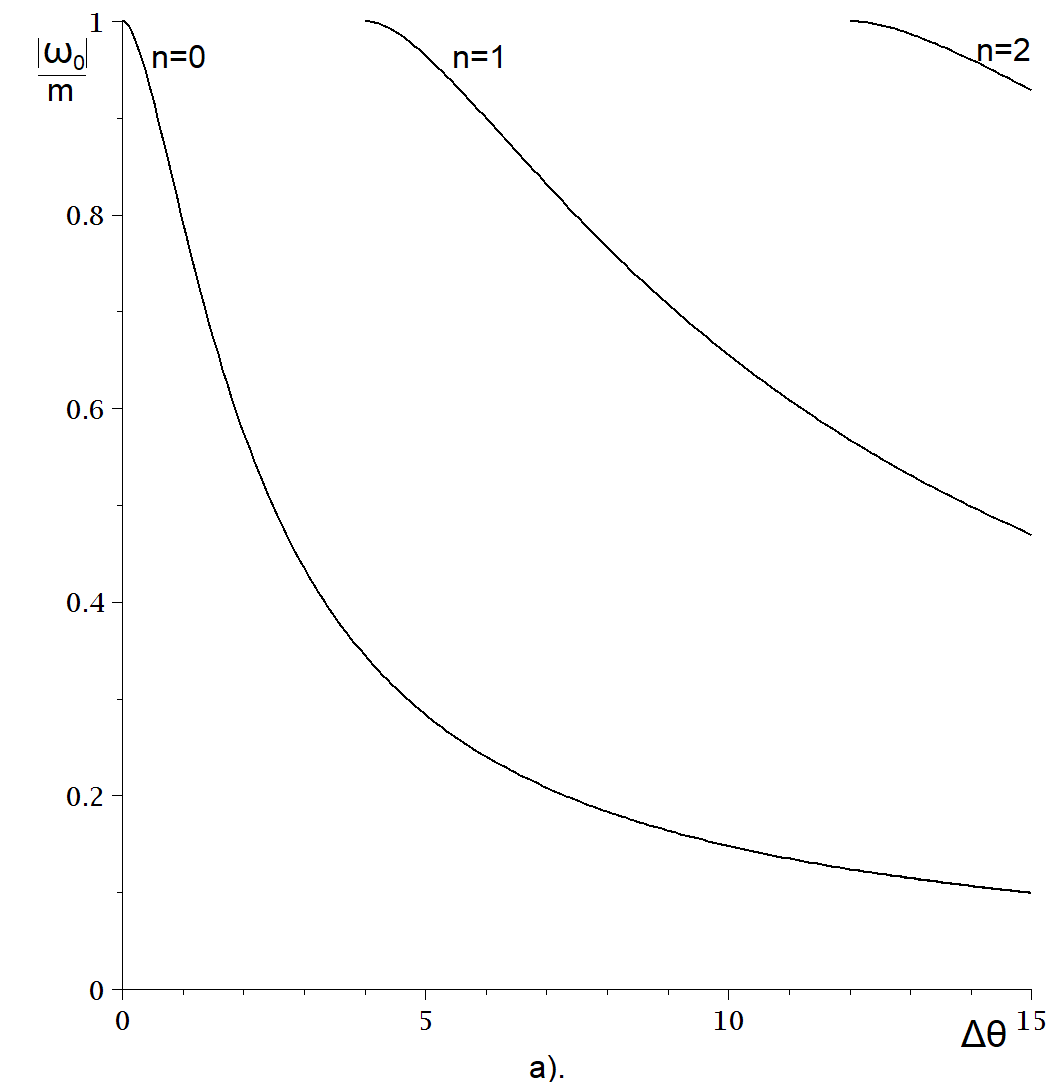}
  \caption{The spectrum of $\omega_0$ for $m/\mu=0.5$}
  \label{fig:spectrum05}
\end{figure}
\begin{figure}[h]
\centering
  \includegraphics[width=0.45\textwidth]{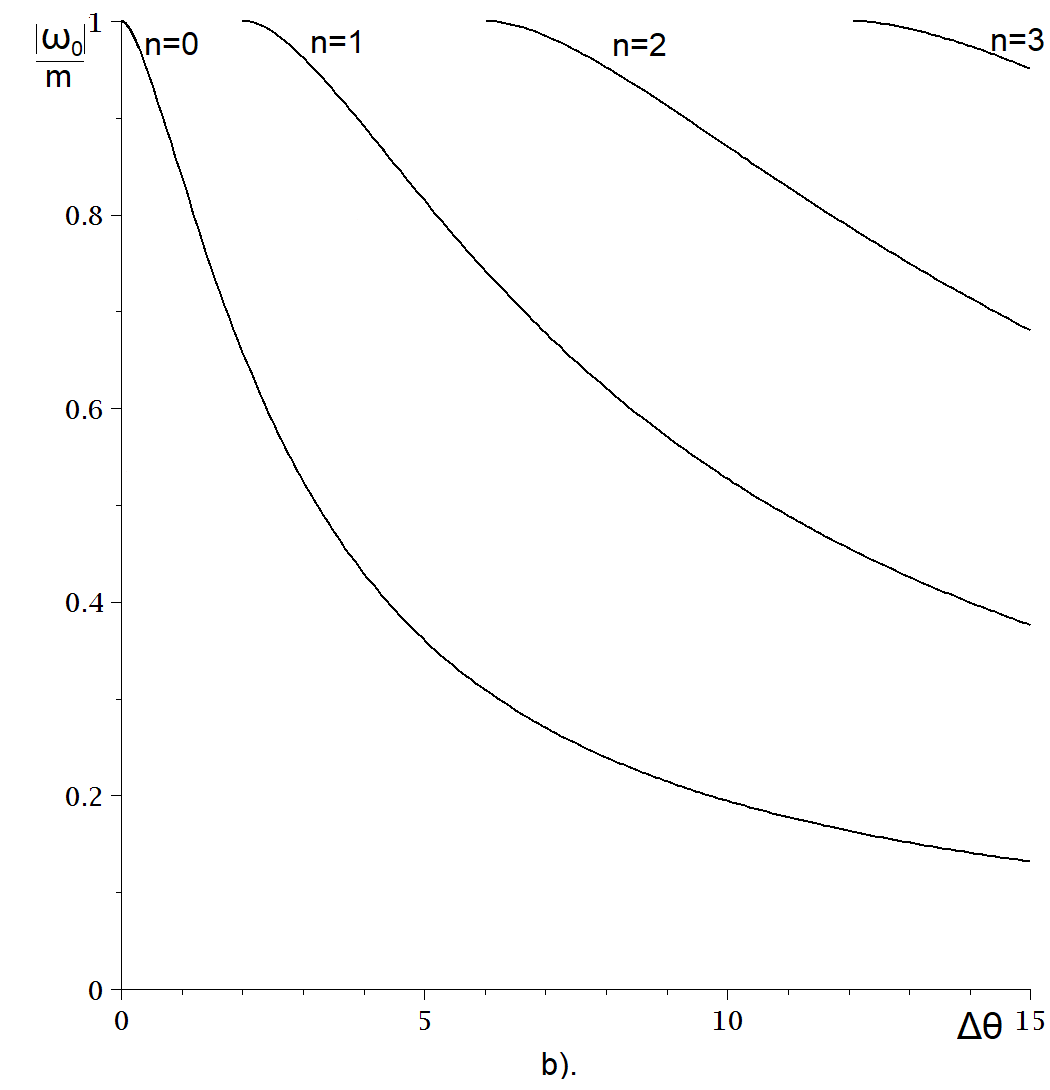}
  \caption{The spectrum of $\omega_0$ for  $m/\mu=1.0$}
  \label{fig:spectrum10}
\end{figure}

The examples of the spectrum for \eqref{ThetaTanh} are presented on Fig. \ref{fig:spectrum05} and Fig. \ref{fig:spectrum10}. Only solutions with $\omega_0<0$ are included, as otherwise the potential introduces a barrier instead of the well and no bound state exist. One can see that the levels appear with growth of the $\Delta\theta$. The phenomena of emergence or disappearance of the levels must lead to the breakdown of the adiabatic approximation and will be explored below.

\section{Correction due to the $\partial_t\theta$ term}

At small times we may approximate the influence of the $\partial_t\theta(t,z)\equiv v(t,z)$ term with the adiabatically changing perturbation that changes the frequencies and profile functions,
\begin{align}
\omega=\omega_0+\omega_1,\quad a=a_0+a_1,\nonumber\\
\omega_1/\omega_0, \ a_1\sim |v|/\omega_0\sim \mu_5/m
\end{align}
The next order equation takes the form,
\begin{align}
\Big[\omega_0^2-\partial_z^2+m^2+\omega(\partial_z\theta)\Big]a_1=iv(\partial_za_0)\nonumber\\
-2\omega_0\omega_1a_0-\omega_1(\partial_z\theta)a_0
\end{align}
Let $a$ be one of the localized bound states we found in the preceding section. For simplicity we will choose it to be real (which can always be done in this dimensionality by the appropriate phase transformation). If we multiply this equation on $a_0^\ast$ and integrate over $z$, l.h.s. vanishes thanks to the Hermiticity of the leading order operator and the equation \eqref{AdiabaticLeading}. From that we derive,
\begin{align}
\omega_1=i\frac{\int_{-\infty}^{+\infty}dz\,va_0\partial_za_0}{\int_{-\infty}^{+\infty}dz\,(2\omega_0+\partial_z\theta)a_0^2}\nonumber\\
=-\frac{i}{2}\frac{\int_{-\infty}^{+\infty}dz\,(\partial_zv)a_0^2}{\int_{-\infty}^{+\infty}dz\,(2\omega_0+\partial_z\theta)a_0^2}
\end{align}
From this it is evident that $\omega_1$ is purely imaginary. This should not be a surprise as the perturbation is non-Hermitian,
\begin{equation}
(iv\partial_z)^\dagger=i\partial_z v=i(\partial_zv)+iv\partial_z
\end{equation}
Therefore the $\partial_t\theta$ term leads to the decay or the amplification of the bound states that were stable in the leading approximation.

Suppose for simplicity that in \eqref{ThetaTanh} only $\bar{\theta}$ and $\Delta\theta$ depend on time. Then for the level $n=0$ we get from \eqref{ZeroMode},
\begin{equation}
\omega_{1,n=0}=-\frac{i}{2}\frac{\frac{\lambda-1}{2\lambda-1}\mu(\partial_t\Delta\theta)}{\omega_0+\frac{\lambda-1}{2\lambda-1}\mu\Delta\theta}
\end{equation}

\section{Schr\"{o}dinger form of the evolution equation}

For various purposes we would like to rewrite our problem as an ordinary Schr\"{o}dinger equation. To do so we introduce,
\begin{equation}
\Psi=\begin{pmatrix}A_{+}\\ i\dot{A}_{+}\end{pmatrix}
\end{equation}
then the equation \eqref{TransverseEq} takes the form,
\begin{equation}
i\partial_t \Psi = \mathcal{H}\Psi,
\label{Schroedingerlike}
\end{equation}
The time-dependent and non-Hermitian Hamiltonian is written as,
\begin{equation}
\mathcal{H}=\begin{pmatrix}0&1\\-\partial_z^2+m^2-i(\partial_t\theta)\partial_z&(\partial_z\theta)\end{pmatrix},
\end{equation}
and satisfies the time-dependent pseudo-Hermiticity relation \cite{Mostafazadeh:2003ps},
\begin{equation}
\rho \mathcal{H}=\mathcal{H}^\dagger\rho-i(\partial_t\rho),
\label{pseudoHermiticity}
\end{equation}
where the matrix $\rho$ is given by,
\begin{equation}
\rho=\begin{pmatrix}-(\partial_z\theta)&1\\1&0\end{pmatrix}
\end{equation}

Therefore we may construct the inner product,
\begin{multline}
\Big(\Psi_1,\Psi_2\Big)=\int_{-\infty}^{+\infty} dz\, \Psi_1^\dagger \rho \Psi_2\\
=\int_{-\infty}^{+\infty}dz\,\Big(iA_{1,+}^\ast\dot{A}_{2,+}-i\dot{A}_{1,+}^\ast A_{2,+}\\
-(\partial_z\theta)A_{1,+}^\ast A_{2,+}\Big),
\label{InnerProduct}
\end{multline}
that is conserved by the Hamiltonian $\mathcal{H}$ despite its non-Hermiticity. Obviously this is a generalization of the ordinary Klein-Gordon inner product and is not positive definite. However, we may restrict our consideration to the positive-norm sector of solutions. On that subspace the equation \eqref{TransverseEq} corresponds to the unitary evolution. The Hamiltonian can be related to the Hermitian Hamiltonian,
\begin{equation}
h=\eta \mathcal{H} \eta^{-1} + i (\partial_t \eta)\eta^{-1},
\end{equation}
where we introduced (restricted to the positive norm subspace),
\begin{equation}
\rho = \eta^\dagger\eta,
\end{equation}
The fact that $\mathcal{H}$ is not similar to $h$ in case of time-dependent $\rho$ means that the spectra of these two operators are not equivalent. Hence, the eigenvalues of $\mathcal{H}$ (that are frequencies $\omega$ considered above) are not guaranteed to be real, and its eigenvectors,
\begin{equation}
\mathcal{H}\Psi_n=\omega_n\Psi_n,
\label{Eigenvector}
\end{equation}
with different $\omega$ are not orthogonal even with respect to the inner product \eqref{InnerProduct},
\begin{equation}
\Big(\Psi_m,\Psi_n)=\frac{i}{\omega_m^\ast-\omega_n}\int_{-\infty}^{+\infty}dz\,\Psi_m^\dagger (\partial_t\rho)\Psi_n,
\label{Nonorthogonal}
\end{equation}
which is obtained from the relation \eqref{pseudoHermiticity}.

\section{Adiabatic regime breakdown}

Using the results from the previous section we may determine the moment of the adiabatic regime breakdown following similar steps to the ordinary Schr\"{o}dinger case.
At every moment we decompose the solution into a superposition of the eigenvectors of $\mathcal{H}$ which will for now consider to be a discrete set,
\begin{equation}
\Psi=\sum_n c_n(t) \Psi_n e^{-i\int dt\,\omega_n}
\end{equation}
Where we assume that each $\Psi_n$ has unit norm with respect to \eqref{InnerProduct}. Then the equation \eqref{Schroedingerlike} results in,
\begin{equation}
\sum_n \Big(\dot{c}_n(t)\Psi_n + c_n(t)\dot{\Psi_n}\Big)\exp\Big(-i\int dt\,\omega_n\Big)=0
\end{equation}
where $\dot{\Psi_n}$ is not an evolution described by \eqref{Schroedingerlike} but a change of $\mathcal{H}$ eigenbasis with time. We are interested when the terms mixing different $c_n$ become significant. For that we multiply this equation on $\Psi_m$ using the inner product \eqref{InnerProduct},
\begin{equation}
\dot{c}_n=-\sum_{m\neq n} \mathcal{N}_{mn} \exp\Big(i\int dt(\omega_m^\ast - \omega_n)\Big)
\end{equation}
We are interested in the mixing terms,
\begin{equation}
\mathcal{N}_{nm}=\dot{c}_n \Big(\Psi_m,\Psi_n\Big) + c_n\Big(\Psi_m,\dot{\Psi}_n\Big)
\end{equation}
In the adiabatic regime $\dot{c}_n\simeq 0$ so we will neglect the first term. Taking a time derivative of \eqref{Eigenvector} and using \eqref{Nonorthogonal} we get,
\begin{multline}
N_{nm}\simeq c_n\Bigg[\frac{i\dot{\omega}}{(\omega_m^\ast-\omega_n)^2}\int_{-\infty}^{+\infty}dz\,\Psi_m^\dagger(\partial_t\rho)\Psi_n\\
-\frac{1}{\omega_m^\ast-\omega_n}\int_{-\infty}^{+\infty}dz\,\Psi_m^\dagger \rho (\partial_t \mathcal{H})\Psi_n
\Bigg]
\label{Mixing}
\end{multline}
As usual, the breakdown of the adiabatic approximation should happen when levels become too close to each other. In other case this happens when $\Delta\theta$ drops sufficiently for the level to approach the continuum spectrum near the point $|\omega|=m$ when the level ceases to exist as can be seen on the Figs. \ref{fig:spectrum05} and \ref{fig:spectrum10}. As can be seen from \eqref{Spectrum} for the $n$-th level that corresponds to,
\begin{equation}
    \Delta\theta\simeq\Delta\theta_{n}=n(n+1)\frac{\mu}{m}
\end{equation}
Let us assume that,
\begin{equation}
\Delta\theta=\Delta\theta_n + \vartheta,\quad \vartheta=-2\mu_5 t
\end{equation}
where $\mu_5$ is the axial chemical potential inside the fireball. Let us define $|\omega|=m-\delta$ and assume that both $\delta$ and $\vartheta$ are small. Then from \eqref{Spectrum} we get,
\begin{equation}
\delta\simeq\frac{1}{2}\frac{m}{(2n+1)^2}\vartheta^2
\end{equation}
Which results in the following estimates,
\begin{align}
\Big|\frac{i\dot{\omega}}{(\omega_m^\ast-\omega_n)^2}\int_{-\infty}^{+\infty}dz\,\Psi_m^\dagger(\partial_t\rho)\Psi_n\Big|\nonumber\\
\sim 16\frac{\mu_5^2 (2n+1)^2}{m\vartheta^3},\\
\Big|\frac{1}{\omega_m^\ast-\omega_n}\int_{-\infty}^{+\infty}dz\,\Psi_m^\dagger \rho (\partial_t \mathcal{H})\Psi_n\Big|\sim \frac{4\mu_5(2n+1)^2}{\vartheta^2}
\end{align}
To break the adiabatic regime one of these contributions must be comparable to the energy i.e. $\sim m$. This gives,
\begin{equation}
\vartheta_{n.a.}\sim \mathrm{max}\Bigg[\Big(\frac{4\mu_5}{m}(2n+1)\Big)^{2/3},(2n+1)\sqrt{\frac{4\mu_5}{m}}\Bigg]
\label{NAestimate}
\end{equation}

\begin{figure}[h]
\centering
  \includegraphics[width=0.45\textwidth]{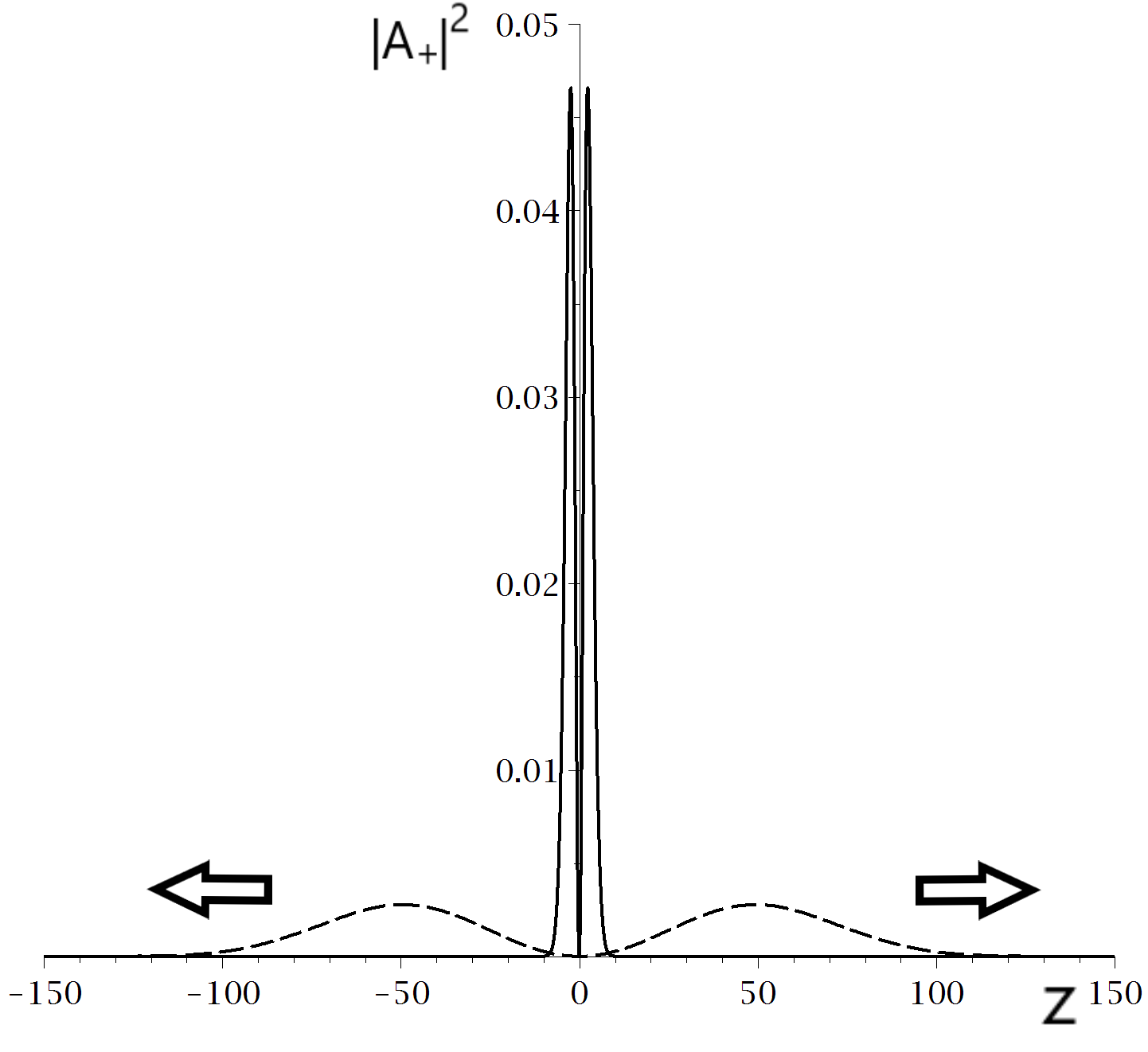}
  \caption{After the breakdown of the adiabatic regime the localized state (solid) transitions into two unbound wavepackets (dashed) travelling towards opposite directions. This particular plot corresponds to $m=3, \mu=1, \mu_5=0.05$}
  \label{fig:wp1}
\end{figure}

\begin{figure}[h]
\centering
  \includegraphics[width=0.45\textwidth]{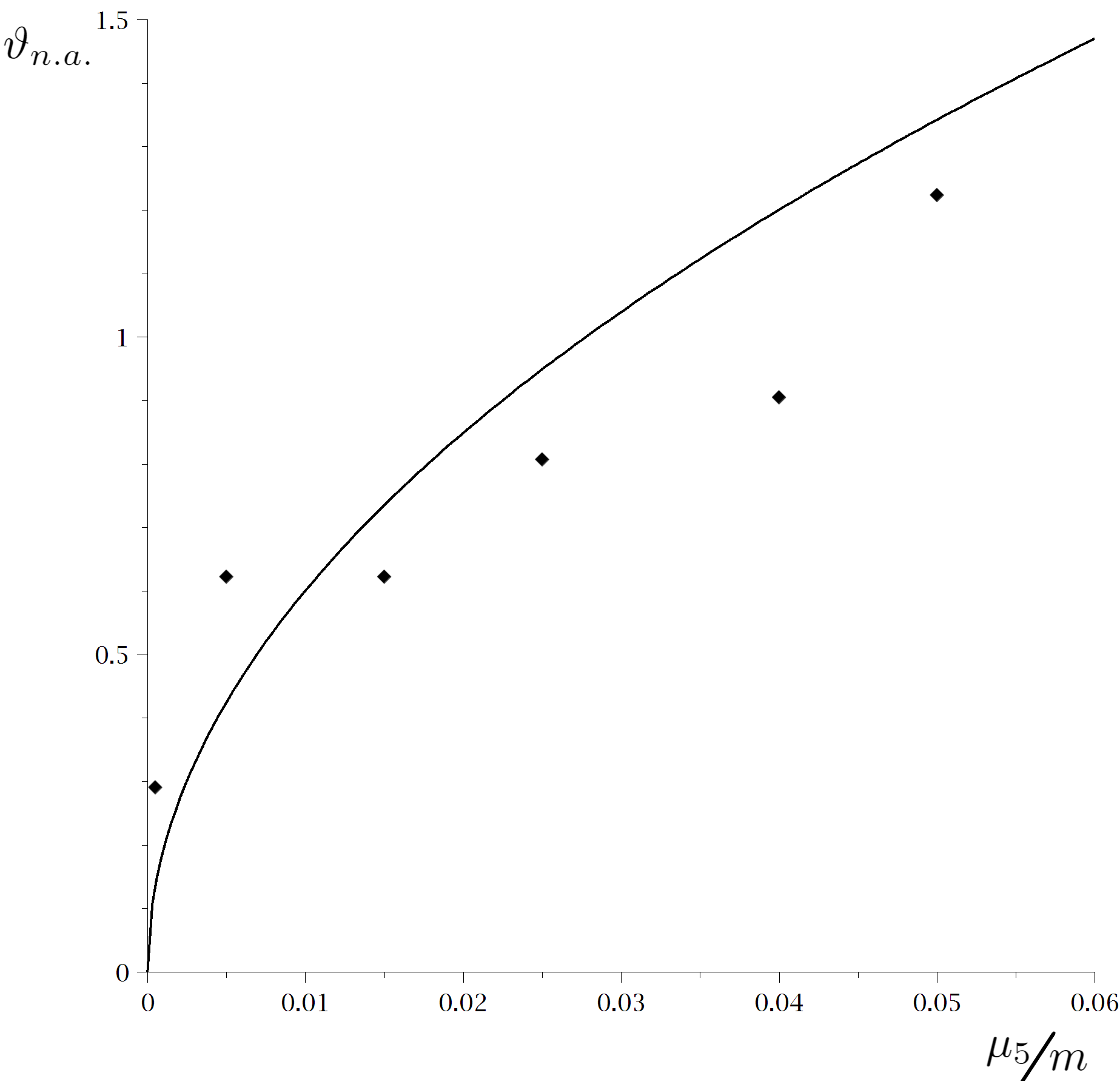}
  \caption{Comparison between adiabatic regime breaking point estimated from the peak of the numerically obtained spectrum and estimate \eqref{NAestimate} for $m=\mu$}
  \label{fig:NA1}
\end{figure}

\begin{figure}[H]
\centering
  \includegraphics[width=0.45\textwidth]{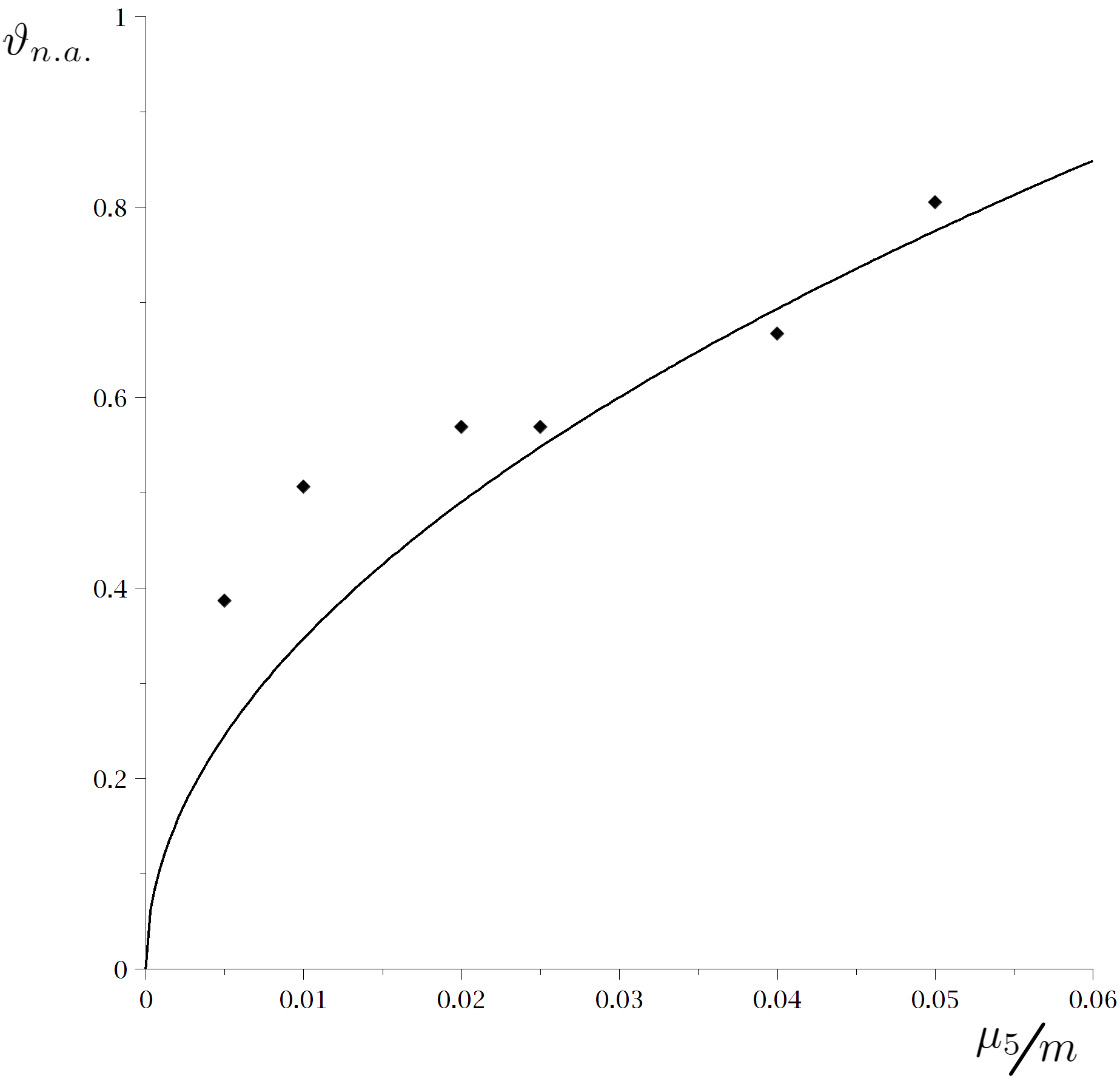}
  \caption{Comparison between adiabatic regime breaking point estimated from the peak of the numerically obtained spectrum and estimate \eqref{NAestimate} for $m=3\mu$}
  \label{fig:NA3}
\end{figure}

The numerical simulations (using second order finite difference on a finite interval of $z$ with Dirichlet boundary conditions) for $n=1$ state with $\Delta\theta=-(2\mu_5 t) h(-t)$, where $h(t)$ is a Heaviside function show what happens after the adiabatic regime breakdown (Fig. \ref{fig:wp1}). During the adiabatic regime as $\Delta\theta$ approaches $\Delta\theta_n$ the bound state profile widens. However, near $\Delta\theta\simeq \Delta\theta_n + \vartheta_{n.a.}$ this widening stops and the bound state transforms into two unbounded wavepackets moving to opposite directions. Their combined momentum representation is close to the momentum representation of the bound state at the adiabatic regime breaking point. Indeed, if we estimate the $\vartheta$ for which the location of the peak of the spectrum for $a_{0,n=1}$ coincides with the numerically obtained one, we get the values close to the estimate \eqref{NAestimate} (see Figs. \ref{fig:NA1} and \ref{fig:NA3})

\section{Conclusions}

We have demonstrated that the massive vector field on a parity-violating background may possess localized states on the boundary of the parity-violating medium. While greatly simplified, this model may imply that a similar statement is valid for the vector meson states in a chiral-breaking medium \cite{Kovalenko:2020ryt}. The time evolution of the background means that the localized states are not exactly stationary. This makes possible for the boundary currents to appear in the transition regions between the domains with different values of the parity-breaking parameter. 

Presently, only vector meson masses $m\simeq 770$ MeV are available from the experimental data whereas only theoretical estimates exist for the value of the axial chemical potential $\mu_5$ or the associated parity-breaking topological charge fluctuations based on different approaches \cite{Kharzeev:2001ev,Lappi:2006fp,Jiang:2016wve,Muller:2016jod,Mace:2016shq,Shi:2017cpu}. If the axial chemical potential at hadronization is as low as $\mu_5\simeq 0.2 T$ \cite{Becattini:2020xbh} with $T\sim 150$ MeV then it is indeed much smaller than $m$ so $\theta$ may be assumed to change slowly. We assume that the spatial parameter $\mu$ also should lie in the range $10 - 1000$ MeV.

Our treatment would be valid only as long as the adiabatic approximation holds. However, as we noted, it may be expected to break during the evolution of the parity-breaking medium leading to the annihilation of the boundary currents into the ingoing and outgoing particles with potentially rich phenomenology. The spectrum of the resulting unbound particles is determined by the bound state profile at the point of the adiabatic regime breakdown, which is determined by the instant $\mu_5$ and $\mu$ values rather than by a history of the background evolution. This suggests that such explosion processes may appear as exotic spikes in the vector meson production.

At this point, however, it is extremely hard to make specific conclusions about experimental implications of the existence and explosions of such boundary currents. For that, one also has to study the interaction between the bound and unbound states, and investigate the possible pumping of the boundary currents by the background. It is also important to study the backreaction of such processes on the parity-breaking background evolution. We employ the generalized chiral perturbation theory and vector meson dominance model \cite{Andrianov:2019fwz,Andrianov:2022rsh,Vioque-Rodriguez:2021hgp,Kovalenko:2020ryt} to explore these questions in \cite{KNZupcoming}.

\section*{Acknowledgements}
We would like to thank Alexander Andrianov for the discussions and suggestions. This research was funded by the Russian Science Foundation grant number 22-22-00493.


\begin{thebibliography}{10}

\bibitem{schwartz2014quantum}
Matthew~D Schwartz.
\newblock {\em Quantum field theory and the standard model}.
\newblock Cambridge University Press, 2014.

\bibitem{chen2021heavy}
Shanzhen Chen, Yiming Li, Wenbin Qian, Yuehong Xie, Zhenwei Yang, Liming Zhang, and Yanxi Zhang.
\newblock Heavy flavour physics and cp violation at lhcb: a ten-year review.
\newblock {\em arXiv preprint arXiv:2111.14360}, 2021.

\bibitem{khriplovich2012cp}
Iosif~B Khriplovich and Steve~K Lamoreaux.
\newblock {\em CP violation without strangeness: electric dipole moments of particles, atoms, and molecules}.
\newblock Springer Science \& Business Media, 2012.

\bibitem{ACME:18}
Vitaly Andreev, DG~Ang, D~DeMille, JM~Doyle, G~Gabrielse, J~Haefner, NR~Hutzler, Z~Lasner, C~Meisenhelder, BR~O'Leary, et~al.
\newblock Improved limit on the electric dipole moment of the electron.
\newblock {\em Nature}, 562(7727):355--360, 2018.

\bibitem{zakharova2022rotating}
Anna Zakharova.
\newblock Rotating and vibrating symmetric-top molecule raoch 3 in fundamental p, t-violation searches.
\newblock {\em Physical Review A}, 105(3):032811, 2022.
\newblock [Phys. Rev. A 105, 069902(E) (2022)].

\bibitem{ourRaOH}
Anna Zakharova and Alexander Petrov.
\newblock {$\mathcal{P}$,$\mathcal{T}$-odd effects for the RaOH molecule in the excited vibrational state}.
\newblock {\em Phys. Rev. A}, 103(3):032819, 2021.

\bibitem{zakharova2021rovibrational}
Anna Zakharova, Igor Kurchavov, and Alexander Petrov.
\newblock Rovibrational structure of the ytterbium monohydroxide molecule and the p, t-violation searches.
\newblock {\em The Journal of Chemical Physics}, 155(16):164301, 2021.

\bibitem{petrov2022sensitivity}
Alexander Petrov and Anna Zakharova.
\newblock Sensitivity of the yboh molecule to p t-odd effects in an external electric field.
\newblock {\em Physical Review A}, 105(5):L050801, 2022.

\bibitem{harris1999new}
P~G Harris, C~A Baker, K~Green, P~Iaydjiev, S~Ivanov, D~J~R May, J~M Pendlebury, D~Shiers, K~F Smith, M~Van~der Grinten, et~al.
\newblock New experimental limit on the electric dipole moment of the neutron.
\newblock {\em Physical review letters}, 82(5):904, 1999.

\bibitem{Cheng1988}
Hai-Yang Cheng.
\newblock The strong {CP} problem revisited.
\newblock {\em Physics Reports}, 158(1):1--89, 1988.

\bibitem{KimCarosi2010}
Jihn~E. Kim and Gianpaolo Carosi.
\newblock {Axions and the Strong CP Problem}.
\newblock {\em Rev. Mod. Phys.}, 82:557--602, 2010.
\newblock [Rev.Mod.Phys. 91, 049902(E) (2019)].

\bibitem{Kharzeev:1998kz}
Dmitri Kharzeev, R.~D. Pisarski, and Michel H.~G. Tytgat.
\newblock {Possibility of spontaneous parity violation in hot QCD}.
\newblock {\em Phys. Rev. Lett.}, 81:512--515, 1998.

\bibitem{buckley2000can}
K~Buckley, T~Fugleberg, and A~Zhitnitsky.
\newblock Can induced $\theta$ vacua be created in heavy-ion collisions?
\newblock {\em Physical Review Letters}, 84(21):4814, 2000.

\bibitem{kharzeev2006parity}
Dmitri Kharzeev.
\newblock Parity violation in hot qcd: Why it can happen, and how to look for it.
\newblock {\em Physics Letters B}, 633(2-3):260--264, 2006.

\bibitem{Andrianov:2007kz}
A.~A. Andrianov and D.~Espriu.
\newblock {On the possibility of P-violation at finite baryon-number densities}.
\newblock {\em Phys. Lett. B}, 663:450--455, 2008.

\bibitem{Andrianov:2009pm}
A.~A. Andrianov, V.~A. Andrianov, and D.~Espriu.
\newblock {Spontaneous P-violation in QCD in extreme conditions}.
\newblock {\em Phys. Lett. B}, 678:416--421, 2009.

\bibitem{kharzeev2008effects}
Dmitri~E Kharzeev, Larry~D McLerran, and Harmen~J Warringa.
\newblock The effects of topological charge change in heavy ion collisions:“event by event p and cp violation”.
\newblock {\em Nuclear Physics A}, 803(3-4):227--253, 2008.

\bibitem{fukushima2008chiral}
Kenji Fukushima, Dmitri~E Kharzeev, and Harmen~J Warringa.
\newblock Chiral magnetic effect.
\newblock {\em Physical Review D}, 78(7):074033, 2008.

\bibitem{buividovich2009numerical}
P~V Buividovich, Maxim~N Chernodub, E~V Luschevskaya, and M~I Polikarpov.
\newblock Numerical evidence of chiral magnetic effect in lattice gauge theory.
\newblock {\em Physical Review D}, 80(5):054503, 2009.

\bibitem{abelev2009azimuthal}
BI~Abelev, MM~Aggarwal, Z~Ahammed, AV~Alakhverdyants, BD~Anderson, D~Arkhipkin, GS~Averichev, J~Balewski, O~Barannikova, LS~Barnby, et~al.
\newblock Azimuthal charged-particle correlations and possible local strong parity violation.
\newblock {\em Physical review letters}, 103(25):251601, 2009.

\bibitem{fukushima2010electric}
Kenji Fukushima, Dmitri~E Kharzeev, and Harmen~J Warringa.
\newblock Electric-current susceptibility and the chiral magnetic effect.
\newblock {\em Nuclear Physics A}, 836(3-4):311--336, 2010.

\bibitem{yamamoto2011lattice}
Arata Yamamoto.
\newblock Lattice study of the chiral magnetic effect in a chirally imbalanced matter.
\newblock {\em Physical Review D}, 84(11):114504, 2011.

\bibitem{Kharzeev:2013ffa}
Dmitri~E. Kharzeev.
\newblock {The Chiral Magnetic Effect and Anomaly-Induced Transport}.
\newblock {\em Prog. Part. Nucl. Phys.}, 75:133--151, 2014.

\bibitem{haque2019measurements}
Md~Rihan Haque, Alice Collaboration, et~al.
\newblock Measurements of the chiral magnetic effect in pb--pb collisions with alice.
\newblock {\em Nuclear Physics A}, 982:543--546, 2019.

\bibitem{agakishiev2008study}
G~Agakishiev, C~Agodi, H~Alvarez-Pol, A~Balanda, R~Bassini, G~Bellia, D~Belver, A~Belyaev, A~Blanco, M~B{\"o}hmer, et~al.
\newblock Study of dielectron production in c+ c collisions at 1agev.
\newblock {\em Physics Letters B}, 663(1-2):43--48, 2008.

\bibitem{lapidus2009low}
KO~Lapidus and VM~Emel’yanov.
\newblock Low mass dilepton production in relativistic heavy ion collisions.
\newblock {\em Physics of Particles and Nuclei}, 40(1):29--48, 2009.

\bibitem{ElectroProbes}
Itzhak Tserruya.
\newblock Electromagnetic probes.
\newblock In R.~Stock, editor, {\em Datasheet from Landolt-B{\"o}rnstein - Group I Elementary Particles, Nuclei and Atoms {\textperiodcentered} Volume 23: ``Relativistic Heavy Ion Physics'' in SpringerMaterials}. Springer-Verlag Berlin Heidelberg, Berlin Heidelberg, 2010.

\bibitem{andrianov2012dilepton}
AA~Andrianov, VA~Andrianov, D~Espriu, and X~Planells.
\newblock Dilepton excess from local parity breaking in baryon matter.
\newblock {\em Physics Letters B}, 710(1):230--235, 2012.

\bibitem{andrianov2014analysis}
Alexander~A Andrianov, Vladimir~A Andrianov, Dom{\`e}nec Espriu, and Xumeu Planells.
\newblock Analysis of dilepton angular distributions in a parity breaking medium.
\newblock {\em Physical Review D}, 90(3):034024, 2014.

\bibitem{Andrianov:2017hbf}
A.~A. Andrianov, V.~A. Andrianov, D.~Espriu, A.~E. Putilova, and A.~V. Iakubovich.
\newblock {Decays of light mesons triggered by chiral chemical potential}.
\newblock {\em Acta Phys. Polon. Supp.}, 10:977, 2017.

\bibitem{Andrianov:2017ilv}
A.~A. Andrianov, V.~A. Andrianov, D.~Espriu, A.~V. Iakubovich, and A.~E. Putilova.
\newblock {Exotic meson decays in the environment with chiral imbalance}.
\newblock {\em EPJ Web Conf.}, 158:03012, 2017.

\bibitem{Andrianov:2016swz}
A.~A. Andrianov, S.~S. Kolevatov, and R.~Soldati.
\newblock {Parity Breaking Medium and Squeeze Operators}.
\newblock {\em Phys. Rev. D}, 95(7):076020, 2017.

\bibitem{Andrianov:2015oxa}
A.~A. Andrianov, V.~A. Andrianov, D.~Espriu, and S.~S. Kolevatov.
\newblock {Stellar matter with pseudoscalar condensates}.
\newblock {\em Eur. Phys. J. C}, 76(3):169, 2016.

\bibitem{Kovalenko:2020ryt}
Vladimir Kovalenko, Alexander Andrianov, and Vladimir Andrianov.
\newblock {Vector mesons spectrum in a medium with a chiral imbalance induced by the vacuum of fermions}.
\newblock {\em J. Phys. Conf. Ser.}, 1690(1):012097, 2020.

\bibitem{Stueckelberg:1957zz}
E.~C.~G. Stueckelberg.
\newblock {Theory of the radiation of photons of small arbitrary mass}.
\newblock {\em Helv. Phys. Acta}, 30:209--215, 1957.

\bibitem{Batalin:1986fm}
I.~A. Batalin and E.~S. Fradkin.
\newblock {Operatorial Quantization of Dynamical Systems Subject to Second Class Constraints}.
\newblock {\em Nucl. Phys. B}, 279:514--528, 1987.

\bibitem{flugge2012practical}
Siegfried Fl{\"u}gge.
\newblock {\em Practical quantum mechanics}.
\newblock Springer Science \& Business Media, 2012.

\bibitem{Mostafazadeh:2003ps}
Ali Mostafazadeh.
\newblock {Quantum mechanics of Klein-Gordon type fields and quantum cosmology}.
\newblock {\em Annals Phys.}, 309:1--48, 2004.

\bibitem{Kharzeev:2001ev}
D.~Kharzeev, A.~Krasnitz, and R.~Venugopalan.
\newblock {Anomalous chirality fluctuations in the initial stage of heavy ion collisions and parity odd bubbles}.
\newblock {\em Phys. Lett. B}, 545:298--306, 2002.

\bibitem{Lappi:2006fp}
T.~Lappi and L.~McLerran.
\newblock {Some features of the glasma}.
\newblock {\em Nucl. Phys. A}, 772:200--212, 2006.

\bibitem{Jiang:2016wve}
Yin Jiang, Shuzhe Shi, Yi~Yin, and Jinfeng Liao.
\newblock {Quantifying the chiral magnetic effect from anomalous-viscous fluid dynamics}.
\newblock {\em Chin. Phys. C}, 42(1):011001, 2018.

\bibitem{Muller:2016jod}
Niklas M\"uller, S\"oren Schlichting, and Sayantan Sharma.
\newblock {Chiral magnetic effect and anomalous transport from real-time lattice simulations}.
\newblock {\em Phys. Rev. Lett.}, 117(14):142301, 2016.

\bibitem{Mace:2016shq}
Mark Mace, Niklas Mueller, S\"oren Schlichting, and Sayantan Sharma.
\newblock {Non-equilibrium study of the Chiral Magnetic Effect from real-time simulations with dynamical fermions}.
\newblock {\em Phys. Rev. D}, 95(3):036023, 2017.

\bibitem{Shi:2017cpu}
Shuzhe Shi, Yin Jiang, Elias Lilleskov, and Jinfeng Liao.
\newblock {Anomalous Chiral Transport in Heavy Ion Collisions from Anomalous-Viscous Fluid Dynamics}.
\newblock {\em Annals Phys.}, 394:50--72, 2018.

\bibitem{Becattini:2020xbh}
F.~Becattini, M.~Buzzegoli, A.~Palermo, and G.~Prokhorov.
\newblock {Polarization as a signature of local parity violation in hot QCD matter}.
\newblock {\em Phys. Lett. B}, 822:136706, 2021.
\newblock [Erratum: Phys.Lett.B 826, 136909 (2022)].

\bibitem{Andrianov:2019fwz}
A.~A. Andrianov, V.~A. Andrianov, and D.~Espriu.
\newblock {Chiral perturbation theory vs. Linear Sigma Model in a chiral imbalance medium}.
\newblock {\em Particles}, 3(1):15--22, 2020.

\bibitem{Andrianov:2022rsh}
V.~A. Andrianov, A.~A. Andrianov, and D.~Espriu.
\newblock {The Chiral Medium in a Generalized Sigma Model and the Chiral Perturbation Theory}.
\newblock {\em Phys. Part. Nucl.}, 53(2):111--116, 2022.

\bibitem{Vioque-Rodriguez:2021hgp}
Andrea Vioque-Rodr\'\i{}guez, Angel~G\'omez Nicola, and Dom\`enec Espriu.
\newblock {Studying chiral imbalance using Chiral Perturbation Theory.}
\newblock {\em PoS}, PANIC2021:369, 2022.

\bibitem{KNZupcoming}
Vladimir Kovalenko, Oleg Novikov, and Anna Zakharova.
\newblock in preparation.

\end{thebibliography}

\end{document}